\documentstyle[preprint,aps]{revtex}
\begin{document}
\draft
\preprint{
\parbox{4cm}{
\baselineskip=12pt
KEK-TH-651\\ 
September, 1999\\
\hspace*{1cm}
}}
\title{On Effective Theory of Brane World with Small Tension}
\author{Junji Hisano
\thanks{e-mail: junji.hisano@kek.jp}
and Nobuchika Okada
\thanks{e-mail: okadan@camry.kek.jp, JSPS Research Fellow}}
\address{Theory Group, KEK, Tsukuba, Ibaraki 305-0801, Japan}
%
\maketitle
\begin{abstract}
The five dimensional theory compactified on $S^1$ with two ``branes'' 
(two domain walls) embedded in it is constructed, based on the field-theoretic 
mechanism to generate the ``brane''. Some light states localized in the 
``brane'' appear in the theory. One is the Nambu-Goldstone boson, 
which corresponds to the breaking of the translational invariance in the 
transverse direction of the ``brane''. In addition, if the tension of 
the ``brane'' is smaller than the fundamental scale of the original theory, 
it is found that there may exist not only massless states but also some 
massive states lighter than the fundamental scale in the ``brane''.  
We analyze the four dimensional effective theory by integrating out the 
freedom of the fifth dimension. We show that some effective couplings can 
be explicitly calculated. As one of our results, some effective couplings of 
the state localized in the ``brane'' to the higher Kaluza-Klein modes in 
the bulk are found to be suppressed by the width of the ``brane''. 
The resultant suppression factor can be quantitatively different from the 
one analyzed by Bando et al. using the Nambu-Goto action, 
while they are qualitatively the same. 
\end{abstract}
\newpage
%
\section{Introduction}
\label{sec:introduction}
Recently, extra-dimensional theories have been intensively investigated.  
One of the crucial difference of the recent theories 
with extra-dimensions from the old one like the Kaluza-Klein theories 
\cite{kaluza} is the existence of the ``brane'' configuration which is
embedded in the whole space-time dimensions. Then, we consider the case in
which some fields (for example, fields in the standard model) live on
the ``brane'' and others (graviton, for example) live in the
bulk. With this selection, which field lives where, a new possibility
to solve the hierarchy problem without supersymmetry or technicolor
was first pointed out in ref.\cite{arkani-hamed}. Other aspects of
these theories have been discussed by many authors; phenomenology of 
the Kaluza-Klein (KK) modes \cite{rizzo}, TeV scale unification \cite{dienes},
new interpretation of the fermion mass hierarchy \cite{martin},
neutrino physics \cite{neutrino}, cosmology and astrophysics
\cite{cosmology-astrophysics}.

Since the existence of the ``brane'' plays the most crucial role in
the recent extra-dimensional theories, it is important to investigate  
how the ``brane'' is generated. Some mechanisms to generate the ``brane'' 
configuration are known in the field-theoretic point of view  
(domain wall)\cite{rubakov} or in the string theory (D-brane) 
\cite{polchinsky}. In the following, let us discuss somewhat generically 
expected possibilities whatever the mechanism is. 

Note that the ``brane'' configuration embedded in the extra-dimensions  
breaks the translational invariance in the transverse direction of 
the ``brane''. If the ``brane'' is spontaneously generated, 
there exits the Nambu-Goldstone (NG) mode in the ``brane''. 
Thus, it is expected that the NG mode appears 
in the four dimensional effective theory. 

In general, the ``brane'', if it is the wall with zero width like D-brane, 
is a fluctuating object in quantum mechanical point of view. 
The width of the fluctuation is naturally given by $M_F/\Lambda^2$,  
where $\Lambda$ and $M_F$ are the ``brane'' tension and the fundamental scale  
of the original theory, respectively. If the ``brane'' tension is smaller 
than the fundamental scale, this effect should be considered 
in the effective theory. In fact, Bando et al. argued on ref.\cite{bando} 
that the recoil effect of the fluctuating brane by the NG boson suppresses 
some effective couplings among matters on the brane and the higher KK modes 
in the bulk, and the suppression factor is found to be 
$\exp(-M_F^2 ~m_{KK}^2~/\Lambda^4)$, where $m_{KK}$ 
is the mass of the KK mode. 

Also, there is a possibility that not only massless states but also 
some massive states are localized in the brane. For example, consider 
the field-theoretic mechanism to generate the domain wall \cite{rubakov}. 
Some fields are localized by the potential wall in the direction of 
the extra-dimensions. If the potential is deep enough and its width 
is larger than the inverse of the fundamental scale, there may exist 
the massive spectrum within the range smaller than the fundamental scale. 
In this case, the localized massive states should be considered 
in the four dimensional effective theory. 
 
In this paper, the five dimensional theory compactified on $S^1$ with 
two domain walls embedded in it is constructed based on a concrete 
field-theoretic mechanism to generate the domain wall\cite{rubakov}. 
Due to the existence of the domain wall, the break down of the translational 
invariance in the direction of the fifth dimension leads to the presence 
of the NG boson in the wall, as far as the gravitational effect is negligible. 
In addition, if the width of the ``brane'' $r_W$ is larger than $1/M_F$, 
we can find that some massive modes whose masses are smaller than 
the fundamental scale may be also localized in the wall. We analyze 
the four dimensional effective theory by integrating out the freedom of 
the fifth dimension. We show that some effective couplings can be explicitly  
calculated. As one of the result, some effective couplings of the state 
localized in the ``brane'' to the higher Kaluza-Klein modes in the bulk 
are found to be suppressed by the width of the ``brane'', and the suppression 
factor is $\exp(-c ~r_W^2~ m_{KK}^2)$ with $c$, the combination of 
the coupling constants. In the result of ref.\cite{bando}, 
$r_W$ corresponds to the width of the brane fluctuation $\sim M_F/\Lambda^2$. 
However, they are quantitatively different. 

In Sec. \ref{sec:set-up}, we construct the five dimensional theory 
compactified on $S^1$ with two domain walls based on the field-theoretic 
mechanism to generate the domain wall. The massless localized state appears 
as the fluctuation mode from the background domain wall configuration. 
This is just the NG boson with respect to the breaking of the translational 
invariance in the direction of the fifth dimension. 
In Sec. \ref{sec:domain-wall-fermions}, we consider the fermion in the bulk  
which couples to the domain wall background. In addition to a localized 
chiral fermion known as the ``domain wall fermion'', the existence 
of the localized massive states is discussed in the case that the wall width  
is larger than the inverse of the fundamental scale. We analyze the four 
dimensional effective theory in Sec. \ref{sec:effective-theory}. 
The effective couplings among some fields in the wall or among fields 
in the walls and the bulk are explicitly calculated. 
Sec. \ref{sec:summary-comments} is devoted to summary and comments. 

%
%
\section{Set-up}
\label{sec:set-up}
First of all, we discuss the mechanism to generate the domain wall 
in five dimensional space-time \cite{rubakov}. Let us consider 
a real scalar field in five dimensions,  
\begin{eqnarray} 
 {\cal L}_{(5)} = 
  \frac{1}{2} \partial_{M} \phi \partial^{M} \phi - V(\phi)   \; ,  
\end{eqnarray}
where 
\begin{eqnarray} 
 V(\phi) = \frac{m_\phi^4}{2 \lambda} 
	   -m_\phi^2 \phi^2  
           + \frac{\lambda}{2} \phi^4 \; .  \end{eqnarray}
Here $\phi= \phi(x,y)$, and $x=x^\mu \; (\mu=0,1,2,3)$ and $y$ 
is the coordinates of the four dimensional space-time and the fifth dimension, 
respectively. Note that $\phi$ and $\lambda$ have mass dimensions $3/2$ 
and $-1$, respectively. It is well known that there is a non-trivial 
background configuration $\phi_0 (y)$ as a solution of the equation of motion, 
called the kink solution, in the direction of the fifth dimension, 
\begin{eqnarray} \phi_0 (y) = \frac{m_\phi}{\sqrt{\lambda}} 
\tanh [m_\phi y] \; .  \label{kink} \end{eqnarray}
Here, we fixed the kink center at $y=0$. The domain wall is generated 
by the vacuum expectation value of $\phi$, and its energy density is given by 
\begin{eqnarray} 
 \rho(y) = \frac{1}{2} \left(\frac{d \phi_0}{d y}\right)^2 + V(\phi_0) 
         = \frac{m_\phi^4}{\lambda} \cosh^{-4} [m_\phi y] \; .   
\end{eqnarray}
We define the ``brane'' tension of the wall as 
\begin{eqnarray} 
 \Lambda^4 = \int_{-\infty}^{\infty} dy \; \rho(y)  
           = \frac{4}{3}  \frac{m_\phi^3}{\lambda}\; .  
 \label{tension} 
\end{eqnarray}
From eq.(\ref{kink}) the width of the ``brane'' $r_W$ is given by $1/m_\phi$. 
In the following discussion, we consider the case that the tension 
of the ``brane'' is smaller than the fundamental scale and/or the width 
of the wall are larger than the inverse of the fundamental scale; 
$m_\phi < M_F$. 

In the WKB approximation, the spectrum of perturbation in the presence 
of the background kink can be found by solving the linearized equation 
of motion for the field $\phi(x, y)= \phi_0(y) + U_{\eta}(y) \eta(x)$. 
For the massless mode $\partial^2_{(4)}\eta(x)=0$, the solution is given by 
\begin{eqnarray} 
  U_{\eta}(y) = \frac{\sqrt{3 m_\phi} }{2}
	\cosh^{-2}[m_\phi y]  \; . 
\end{eqnarray}
The kinetic term of $\eta(x)$ is canonically normalized by $U_\eta (y)$. 
This massless mode corresponds to the freedom of the shift of the kink 
solution in the direction of the fifth dimension, 
$\phi_0(y) \rightarrow \phi_0(y + \eta(x))$. 
Once we fix the kink center at a point, the translational invariance 
in the direction of the fifth dimension is spontaneously broken. 
Therefore, the massless mode is the NG mode with respect to this break down, 
and the decay constant is given by $d \phi_0 /dy \propto U_\eta(y)$. 
Also, there is one massive mode $ \tilde{\eta}(x)$ localized 
around the kink center with mass squared $m^2 = 3 m_\phi^2$ \cite{dashen}. 
For this state, the solution of the linearized equation is given by 
\begin{eqnarray}
 U_{\tilde{\eta}}(y) = \sqrt{ \frac{3}{2} m_\phi} 
               \sinh[m_\phi y] \cosh^{-2} [m_\phi y] \; .
\end{eqnarray}
The kinetic term of $\tilde{\eta}(x)$ is canonically normalized 
by $U_{\tilde{\eta}}(y)$. 

Up to now, the compactification of the fifth dimension is not discussed. 
In the following discussion, we consider the five dimensional theory 
compactified on the $S^1$ with radius $R$, in which the domain wall is 
embedded. In this case, the kink solution discussed above is not a solution 
of the compactified theory, because it cannot satisfy the periodic boundary 
condition $\phi_0(y) = \phi_0 (y +2 \pi R)$. In order to get such a solution, 
we introduce the anti-kink solution centered at $y= \pi R$ and connect it 
to the kink solution with the dilute gas approximation. 
The approximate solution is given by 
\begin{eqnarray} 
  \phi_0(y) = \frac{m_\phi}{\sqrt{\lambda}} 
	\tanh [m_\phi y] \tanh [m_\phi (\pi R- y)] \; , 
\end{eqnarray}
with the periodic boundary condition $\phi_0(y)=\phi_0(y+2 \pi R)$. 
In order for this approximation to be justified, 
$r_W (\equiv1/m_\phi) \ll \pi R$ is assumed. Since the distance between 
the kink center and the anti-kink center is very large compared with 
the width of each domain wall, the overlapping of two domain walls can be 
neglected. In the paper, we assume that the domain wall centered at $y=0$ 
is the ``brane'' in which we are living, and concentrate on only the domain 
wall around the kink center. In our case, the solution of eq.(\ref{kink}) 
can be regarded as a good approximation. We do not have to pay attention 
to the existence of the other domain wall centered at $y= \pi R$ 
in the following. Now, the compactified five dimensional theory 
in which two domain walls  are embedded has been constructed.
\footnote{
This configuration where a pair of kink and anti-kink exists 
 on the compact space is not stable for the small fluctuation. 
However, it is a difficult problem to construct a stable configuration 
 with some kink solutions on a compact space, and it is out of our scope. 
Here we assume the existence of the mechanism of the stabilization.} 
%
\section{Domain Wall Fermions}
\label{sec:domain-wall-fermions} 
In this section, consider the bulk fermion which has a Yukawa coupling to 
the background kink solution. It is well known that there is a localized 
chiral fermion in the domain wall \cite{rubakov}. If the width of 
the ``brane'' is larger than the inverse of the fundamental scale, we can 
find that not only the chiral fermion but also massive states lighter than 
the fundamental scale may be localized in the wall. 

Let us consider the the bulk fermion Lagrangian of the form, 
\begin{eqnarray}
 {\cal L}_{(5)} = \bar{\Psi}(x,y) 
  \left( i \Gamma^M \partial_M + g_Y \phi_0(y)\right)  \Psi(x,y) \; , 
 \label{yukawa}
\end{eqnarray}
where $\Psi$ is a four component spinor in five dimensions, and 
the five-dimensional $\gamma$-matrices are 
$\Gamma^\mu = \gamma^\mu$ for $\mu = 0,\ldots,3$ and 
$\Gamma^4 = i \gamma_5$, respectively. Here, we assume that $g_Y$ is real 
and positive. Suppose that the fermion $\Psi$ is described by Weyl fermions 
in four dimensional space-time such that  
\begin{eqnarray}
 \Psi(x,y) = U_L(y) \psi_L(x) + U_R(y) \psi_R(x)   \; . 
\end{eqnarray}
The above Lagrangian can be rewritten as 
\begin{eqnarray}
 {\cal L}_{(5)} &=& |U_L|^2 \overline{\psi_L} i \gamma^\mu \partial_\mu \psi_L
   +  |U_R|^2 \overline{\psi_R} i \gamma^\mu \partial_\mu \psi_R \\ \nonumber 
  &+& \overline{\psi_L} U_L^\dagger 
    \left( - U_R^\prime  + U_R \; g_Y \phi_0 \right) \psi_R \\ \nonumber
 &+& \overline{\psi_R}  U_R^\dagger
    \left( + U_L^\prime  + U_L \; g_Y \phi_0 \right) \psi_L \; ,  
\end{eqnarray}
where prime denotes the derivative with respect to $y$. 
Considering the Dirac equation in four dimensions, 
$i \gamma^\mu \partial_\mu \psi_{L,R} = m \psi_{R,L}$, 
the equations of motion for $U_{L,R}$ are given by 
\begin{eqnarray}
  U_L^{\prime \prime} &+& \left
  ( m^2 + g_Y \phi_0^\prime - g_Y^2 \phi_0^2 \right) U_L =0  \; ,\\ 
  U_R^{\prime \prime} &+& \left
  ( m^2 - g_Y \phi_0^\prime - g_Y^2 \phi_0^2 \right) U_R =0 \; . 
\end{eqnarray}
It is useful to rewrite these equations into the form 
\begin{eqnarray} 
  U_L^{\prime \prime} 
   &+& \left( E + \frac{V_L} {\cosh^2[m_\phi y]} 
   \right) U_L =0    \label{eqUL}  \; ,  \\ 
  U_R^{\prime \prime}
   &+& \left( E + \frac{V_R} {\cosh^2[m_\phi y]}
   \right) U_R =0   \; ,  \label{eqUR}
\end{eqnarray}
 where 
\begin{eqnarray}
 E &=& m^2 - \frac{g_Y^2}{\lambda} m_\phi^2 \; , \\ 
 V_L &=&  \frac{g_Y}{\sqrt{\lambda}} m_\phi^2 
 (\frac{g_Y}{\sqrt{\lambda}}+1) \; ,\\ 
 V_R &=&  \frac{g_Y}{\sqrt{\lambda}} m_\phi^2 
 (\frac{g_Y}{\sqrt{\lambda}}-1) \; . 
\end{eqnarray}
Note that these equations can be regarded as (1+1)-dimensional 
Schr\"odinger equations with the potential, $ - V_{L,R} \cosh^{-2}[m_\phi y]$.

We are interested in the bound state which is the solution satisfying 
two boundary conditions; $U_{L,R}(0)$ is finite and 
$U_{L,R}(y) \rightarrow 0$ for $|y| \rightarrow \infty$. 
Such a solution is given by using the hyper-geometric function 
$F[a, b; c; z]$. For $U_L(y)$, we find the following mass eigenvalues, 
\begin{eqnarray}
 m^2 = m_\phi^2 n_L \left( 
  2 \frac{g_Y}{\sqrt{\lambda}}-n_L \right) 
 \; \; ; n_L = 0,1,2, \ldots  < g_Y/ \sqrt{\lambda} \; . 
\end{eqnarray}
The eigenstates for even numbers of 
$n_L= 2 n_L^\prime \; (n_L^\prime =0,1,2, \ldots )$ and odd numbers of 
$n_L= 2 n_L^{\prime \prime}+1 \; (n_L^{\prime \prime} =0,1,2,\ldots )$ 
are given by (up to normalization factor) 
\begin{eqnarray} 
 U_L^{n_L^\prime}(y) = \left(
      \cosh[m_\phi y] 
      \right)^{- \frac{g_Y}{\sqrt{\lambda}}} \;
         F[-n_L^\prime , -\frac{g_Y}{\sqrt{\lambda}}+ n_L^\prime
            ; 1/2 ; 1- \cosh^2[m_\phi y]] 
\end{eqnarray}
and 
\begin{eqnarray}
  U_L^{n_L^{\prime \prime}}(y) &=&  \sinh[m_\phi y] 
      \left( \cosh[m_\phi y]
    \right)^{- \frac{g_Y}{\sqrt{\lambda}}} \\ \nonumber
  & \times & 
  F[-n_L^{\prime \prime}, - \frac{g_Y}{\sqrt{\lambda}}
    + n_L^{\prime \prime}+1 ; 3/2 ; 
    1- \cosh^2[m_\phi y]] \; ,  
\end{eqnarray}
respectively. Note that, if $g_Y/\sqrt{\lambda}> 1$, there exists 
at least one massive bound state localized around the kink center. 
The mass eigenvalues are smaller than the fundamental scale with 
our assumption $m_\phi \ll M_F$, and, therefore, such states should 
be considered in the effective theory. 

The solution for $U_R(y)$ can be found by the same manner. 
The mass eigenvalue for $U_R(y)$ is given by 
replacing $n_L$ to $n_R+1$ in the results for $U_L$,  
\begin{eqnarray}
 m^2 = m_\phi^2 (n_R+1) 
  \left( 2 \frac{g_Y}{\sqrt{\lambda}}-(n_R+1) \right) 
  \; \;  ;n_R = 0,1,2,\ldots  < g_Y/ \sqrt{\lambda}-1 \; . 
\end{eqnarray}
The eigenstates for even numbers of 
$n_R= 2 n_R^\prime \; (n_R^\prime =0,1,2, \ldots)$ and odd numbers of 
$n_R= 2 n_R^{\prime \prime}+1 \; (n_R^{\prime \prime} =0,1,2,\ldots )$ 
are given by (up to normalization factor) 
\begin{eqnarray} 
  U_R^{n_R^\prime}(y) &=& \left(
      \cosh[m_\phi y]
      \right)^{- \frac{g_Y}{\sqrt{\lambda}}+1} \\ \nonumber
      & \times &  F[-n_R^\prime , 
     - \frac{g_Y}{\sqrt{\lambda}}+ n_R^\prime +1 
            ; 1/2 ; 1- \cosh^2[m_\phi y]] 
\end{eqnarray}
and 
\begin{eqnarray}
  U_R^{n_R^{\prime \prime}}(y) &=& \sinh[m_\phi y]
       \left(  \cosh[m_\phi y]
       \right)^{- \frac{g_Y}{\sqrt{\lambda}}+1} \\ \nonumber
   &\times & 
     F[-n_R^{\prime \prime}, -\frac{g_Y}{\sqrt{\lambda}}
    + n_R^{\prime \prime}+2 ; 3/2 ; 
    1- \cosh^2[m_\phi y]] \; , 
\end{eqnarray}
respectively. Dirac fermions with masses 
$m_n^2 = m_\phi^2 \; n \; (2 g_Y/ \sqrt{\lambda}-n)\; ( n=1,2,\ldots )$ 
are composed by the $n$-th left-handed and the $(n-1)$-th right-handed 
fermions, ${\cal L}_{mass}= - m_n \overline{\psi_L^n} \psi_R^{n-1} $. 

Note that the case $n_L=0$ corresponds to a massless state, 
and the left-handed chiral fermion is localized around the kink center, 
\begin{eqnarray}
 \Psi^{n=0}(x,y) = U_L^{n_L=0}(y) \psi_L^0(x) \; , 
\end{eqnarray}
where
\begin{eqnarray}
     U_L^{n_L=0}(y) =  \left(\frac{ m_\phi 
    \Gamma(\frac{g_Y}{\sqrt{\lambda}} + \frac{1}{2})}
   {\sqrt{\pi} \Gamma(\frac{g_Y}{\sqrt{\lambda}})} \right)^{1/2} 
   \left( \cosh[m_\phi y]
    \right)^{- \frac{g_Y}{\sqrt{\lambda}}} \; .
\end{eqnarray}
On the other hand, there is no solution for the right-handed chiral 
fermion which satisfies the boundary conditions. In fact, the right-handed 
chiral fermion is expected to be localized around the anti-kink center, 
$y= \pi R$. The massive bound states are also localized there. 
These eigenstates can be given from the above result 
by exchanging the chirality $L \leftrightarrow R$. 

Since the fifth dimension is compactified on $S^1$, there are KK modes 
of the fermion which are not localized. Unfortunately, we cannot exactly 
solve the eqs.(\ref{eqUL}) and (\ref{eqUR}) without the second boundary 
condition. However, if $E \gg V_{L, R}$, the potential terms can be neglected, 
and these equations is reduced to the usual KK mode equations. 
In this case, the bulk fermion can be expanded by the series of the KK modes, 
\begin{eqnarray}
 \Psi_{L,R}(x,y) \sim  \frac{1}{\sqrt{2 \pi R}}\sum_n  \psi^n_{L,R}(x)
  \;  e^{i \frac{n}{R} y} \; .
\end{eqnarray}
The masses of the KK modes are given by 
$\tilde{m}_n^2 \simeq (n/R)^2 + g_Y^2m_\phi^2/\lambda$ with integer $n$. 

\section{Effective Theory in Four dimensions}
\label{sec:effective-theory}
In the previous sections, all states which we are interested in were given. 
Now, let us analyze the four dimensional effective theory by integrating out 
the freedom of the fifth dimension. We first consider the effective coupling 
among the domain wall fermions and the KK modes in the bulk. 
Suppose that the domain wall fermion has the gauge charge and the gauge boson 
lives in the bulk. The gauge boson is described by the series of expansion 
of the KK modes as, 
\begin{eqnarray}
 A_\mu (x,y) = \frac{1}{\sqrt{2 \pi R}}\sum_n A_\mu^n(x)
  \; e^{i \frac{n}{R} y} 
\end{eqnarray}
with the mass squared $m_{KK}^2 = (n/R)^2$. 
 
The interaction among two domain wall chiral fermions 
and the bulk gauge boson is of the form, 
\begin{eqnarray}
 {\cal L}_{int}^{(5)} = g \;
 \overline{\Psi_L^0} (x,y)  \gamma^M A_M (x,y) \Psi_L^0 (x,y) \; ,
\end{eqnarray}
where $g$ is the gauge coupling constant with mass dimension $-1/2$. 
The effective gauge coupling in the four dimensional theory, defined as  
\begin{eqnarray} 
  {\cal L}_{int}^{(4)} = \sum_n g_n^{eff} \; \overline{\psi_L^0} (x)
                \gamma^\mu A_\mu (x,y) \psi_L^0 (x) \; , 
\end{eqnarray}
is given by 
\begin{eqnarray}
 g_n^{eff} &=& \frac{g}{\sqrt{2 \pi R}} 
   \int_{-\pi R}^{\pi R} dy \;  |U_L^0(y)|^2 \; 
      e^{i \frac{n}{R} y} \\ \nonumber 
    & \sim & \frac{g}{\sqrt{2 \pi R}}  \left( 
 \frac{m_\phi
   \Gamma(\frac{g_Y}{\sqrt{\lambda}} +\frac{1}{2})}
     {\sqrt{\pi} \Gamma(\frac{g_Y}{\sqrt{\lambda}})} \right)
 \int_{-\infty}^{\infty} dy \;  
  \left( \cosh[m_\phi y]
   \right)^{-2 \frac{g_Y}{\sqrt{\lambda}}} \; e^{i \frac{n}{R} y} \; .   
\end{eqnarray}
Here we assumed $ (g_Y/\sqrt{\lambda}) m_\phi  \gg 1/(\pi R)$, 
because of the asymptotic behavior 
$ \left( \cosh[m_\phi y]  \right)^{-2 g_Y /\sqrt{\lambda}}  
  \sim \exp [-2 (g_Y/\sqrt{\lambda}) m_\phi |y|]$ for large $|y| \gg 1$. 
It is useful to define the dimensionless coupling constant 
$\bar{g} \equiv g/\sqrt{2 \pi R}$, 
since we get $g_n^{eff} = g/\sqrt{2 \pi R}$ for $n=0$. 

After integration with respect to the fifth dimension $y$, 
we get the effective coupling in the four dimensional theory, 
which is very well fitted by the function,  
\begin{eqnarray} 
  g_n^{eff} & \sim & \bar{g} \exp 
   \left[ - \frac14  \frac{\sqrt{\lambda}}{g_Y} 
          \left(\frac{n}{ m_\phi R} \right)^2   \right]  \; . 
\end{eqnarray}
Note that the effective coupling is suppressed for the higher KK modes 
with large $n$. This means the restoration of the translational invariance 
in the direction of the fifth dimension for such KK modes. 

The KK modes with mass $n/R$ much smaller than the inverse of the domain 
width $1/r_W(\sim m_\phi)$ feel the wall thin and hard, and the momentum 
conservation in the direction of the fifth dimension is hardly broken. 
As a result, the couplings become almost universal. On the other hand, 
the modes with mass much larger than $1/r_W$ feel the wall thick and soft, 
and are not aware of the localization of the matter in the domain wall. 
Therefore, the breaking of the momentum conservation becomes restored 
as $n$ becomes large. 

Bando et al. \cite{bando} discussed the effective gauge coupling 
between the KK mode of the gauge boson in bulk and the matter localized 
in the ``brane'', by using the induced metric based on the Nambu-Goto 
action \cite{sundrum}. They showed the effective gauge coupling takes 
of the form as  
\begin{eqnarray} 
  g_n^{eff} &=& \bar{g} \exp\left[ - \frac12
  \left(\frac{n}{R}\right)^2 \frac{M_F^2}{\Lambda^4} \right] \; .
\end{eqnarray}
This suppression factor comes from the quadratically divergent radiative 
correction through the NG boson loops to this vertex. Our result is 
qualitatively consistent with this result, since the quantum fluctuation 
width of the the brane $\sim {M_F}/{\Lambda^2}$ in the Nambu-Goto action 
corresponds to the width of the wall. However, our result is quantitatively 
different from theirs. For example, when $g_Y^2\sim \lambda\sim 1/M_F$, 
the effective coupling becomes 
\begin{eqnarray}
  g_n^{eff} =\bar{g} \exp\left[ - \frac{4}{9} \left(\frac{n}{R}\right)^2 
            \frac{M_F^{2/3}}{\Lambda^{8/3}} \right]  \;  .
\end{eqnarray} 

Next, let us consider the interactions among the localized states in the wall. 
Since these interactions are not suppressed by a volume factor $r_W/R$ 
and are related to the mechanism to generate the ``brane'', they are 
interesting from the phenomenological point of view. As one of examples, 
let us  discuss the Yukawa interaction of the chiral fermion to the massive 
states and the NG boson in the Lagrangian of eq.(\ref{yukawa}), 
\begin{eqnarray}
 {\cal L}_{int}^{(5)} &=& g_Y
     \left( U_\eta(y) \eta(x) \right) 
      \overline{{\Psi}_R^{n_R}}(x,y) {\Psi}^0_L(x,y) \\ \nonumber 
   &=& g_Y 
        \left(  U_\eta(y) U_L^{0 \dagger}(y) U_R^{n_R}(y) \right)
       \; \eta(x)\overline{\psi_R^{n_R}}(x) {\psi}^0_L(x)  \; .  
\end{eqnarray}
Then the effective Yukawa coupling in four dimensions is defined as 
\begin{eqnarray} 
   g_Y^{eff}(n_R) = g_Y
      \int_{-\infty}^{\infty} dy \;
     U_\eta(y) U_L^{0 \dagger}(y) U_R^{n_R}(y) \; .
\end{eqnarray}
Note that $ U_\eta(y)$ and $U_L^0(y)$ are even functions of $y$, 
the effective coupling remains non-zero only in the case that $n_R$ takes 
even numbers. As already discussed, 
for $n_R = 2 n_R^\prime \; (n_R^\prime=0,1,2,\ldots)$, 
$U_R^{n_R^\prime}(y)$ is given by 
\begin{eqnarray}
  U_R^{n_R^\prime} &=&  N_R^{n_R^\prime} 
  \left(  \cosh[m_\phi y] 
  \right)^{-\frac{g_Y}{\sqrt{\lambda}}+1}   \\ \nonumber
 &\times  &  F[ -n_R^\prime, - \frac{g_Y}{\sqrt{\lambda}}
  +n_R^\prime+1  ; 1/2; 
  1-\cosh^2[m_\phi y]] \; , 
\end{eqnarray}
where $N_R^{n_R^\prime}$ is the normalization factor by which 
the kinetic term of $\psi_R^{n_R^\prime}$ is canonically normalized. 
The normalization factor is given by (up to phase) 
\begin{eqnarray}
  N_R^{n_R^\prime} &=&  
    m_\phi^{1/2} \\ \nonumber 
 & \times &
  \left(  \int_{-\infty}^{\infty} dy \; 
  (\cosh y )^{- 2 \left( \frac{g_Y}{\sqrt{\lambda}}-1 \right)} 
   F[ -n_R^\prime, -\frac{g_Y}{\sqrt{\lambda}}+n_R^\prime+1
   ; 1/2; 1-\cosh^2 y]   \right)^{-1/2} \; .
\end{eqnarray}
The explicit description of the effective Yukawa coupling is of the form
\begin{eqnarray}
  g_Y^{eff}(n_R^\prime) 
      = \sqrt{\frac{3}{4}}\;  g_Y \; m_\phi^{1/2} \; 
       W (n_R^\prime,r \equiv \frac{g_Y}{\sqrt{\lambda}}) \; , 
\end{eqnarray}
where 
\begin{eqnarray}
 W \left(n_R^\prime, r \right) 
 &=& \left( \frac{\Gamma(r+1/2)}{\sqrt{\pi} \Gamma(r)} \right)^{1/2} 
 \\ \nonumber
 &\times &\int_{-\infty}^{\infty} dy \;  (\cosh y)^{-2 r -1} 
     F[ -n_R^\prime, -r + n_R^\prime +1 ; 1/2; 1-\cosh^2 y ] \\ \nonumber 
 &\times & \left( 
   \int_{-\infty}^{\infty}dy \;  (\cosh y)^{-2 (r -1)}
   F[ -n_R^\prime, -r + n_R^\prime + 1
   ; 1/2; 1-\cosh^2 y]   \right)^{-1/2}  \; .
\end{eqnarray}
By numerical calculations, we can find that the effective coupling 
for the lowest state $n_R^\prime =0 $ is the largest. For example, 
$W \left(n_R^\prime=0, r \right) \sim 1$,  
$W \left(n_R^\prime=1, r \right) \sim 0.1$ 
and $W \left(n_R^\prime \geq 2, r \right) \ll 0.1$ for $r \gg 1$. 
In this case, $g_Y^{eff}(n_R^\prime=0) \sim 0.3$ 
for $m_\phi /M_F =0.1$ and $g_Y = 1/ \sqrt{M_F}$. 

We do not address the explicit results of other effective couplings. 
One can easily calculate them by using the formulae given in the paper. 
To do this, rewrite the original Lagrangian by using the fields 
in four dimensions and integrate out with respect to the fifth dimension. 
The result depends on the shape of the wave function in the direction of 
the fifth dimension. For example, the couplings of the higher massive bound 
states to the higher KK modes are more suppressed than that of the chiral 
fermion, since the width of the wall becomes larger 
for the higher bound states. 

%
%
\section{summary and comments}
\label{sec:summary-comments}
In summary, we have constructed the five dimensional theory, in which 
the domain wall is embedded, based on the concrete field-theoretic 
mechanism to generate the domain wall. Due to the existence of the domain 
wall and the break down of the translational invariance in the direction 
of the fifth dimension, the NG boson appears in the wall. 
If the tension of the wall is smaller than the fundamental scale, there exist 
not only the massless states but also the massive bound states in the wall. 
They have the masses smaller than the fundamental scale, and appear 
to be considered in the effective theory. 

We have analyzed the four dimensional effective theory by integrating 
out the freedom of the fifth dimension. Some effective couplings have 
been explicitly calculated. We find that the effective couplings 
among two localized states and the higher KK modes, whose masses $n/R$ 
are larger than the inverse of the wall width $1/r_W$, are exponentially 
suppressed as $\exp(-c (n r_W/R)^2)$, so that the breaking of the 
translational invariance becomes restored for such modes. The effective 
Yukawa couplings among one massless state, one massive bound state 
and the NG boson are also calculated, and it is found that the effective 
Yukawa coupling including the lightest massive state is the largest one. 
One can easily calculate other effective couplings 
by using the formulae given in this paper. 

In the phenomenological point of view, the suppression of the effective 
coupling including the KK modes is unfortunate, since this 
suppression makes the effect of the Kaluza-Klein modes in experiments 
loose as discussed in ref.\cite{bando}.  However, there is the Yukawa 
coupling including the NG boson in our effective theory. 
There is a process, for example, 
$\overline{\psi_L^0}\; \psi_L^0 \rightarrow \eta \; \eta$ 
intermediated by the bound state fermions. Such a process may become 
important in a search for the existence of the extra-dimension 
and the domain wall, or in cosmology and astrophysics. 

Our discussion in the five dimensional theory can be extended to more higher 
dimensional theories. For example, in the six-dimensional theory, we can use 
the vortex solution \cite{nielsen} to generate the domain wall. The same 
features of the effective theory discussed above can be expected, 
if the tension of the wall is smaller than the fundamental scale. 

Finally, we would like to give some comments. Our theory discussed above 
has not been gauged, in other words, gravity has not been discussed. 
If we include the gravitational effect in our discussion, the NG boson 
is expected to be eaten up by gauge fields, some components of graviton 
in five dimensions, and to disappear in the effective theory. 
This fact was briefly discussed in ref. \cite{arkani-hamed2} 
considering the fluctuation from the flat background metric, 
and they claimed that $g_{\mu 5}$ components eat the NG boson and 
get masses such that $m_g^2 \sim \Lambda^4/ M_{pl}^2$, 
where $M_{pl}$ is the Planck mass. 
However, note that it is a highly non-trivial problem whether we can find 
a solution of the five-dimensional universe including some domain walls. 
If the existence of the domain wall crucially affects the behavior 
of the background metric, the metric no longer becomes flat 
(as such an example, see ref. \cite{randall}).  
In this case, it seems to be not so easy to understand which field 
is eaten up without a concrete solution. 

Let us see the effect of the gravity for the five dimensional theory 
in which the domain walls are embedded. In this set-up, suppose that 
there exists a solution of the Einstein equation with the metric of the form, 
\begin{eqnarray} 
 ds^2 = \sigma(y) \eta_{\mu \nu} dx^\mu dx^\nu - dy^2 \; ,   
\end{eqnarray} 
by which the Poincare invariance in the four dimensions is ensured. 
Here the metric $\eta_{\mu \nu}= diag(1,-1,-1,-1)$ is used. 
The $(\mu, \nu)$-component of the Einstein equation around the domain wall 
(we fix the center of the domain wall at $y=0$) is found to be 
\begin{eqnarray}
  \frac{d^2 \sigma(y)}{d y^2} \Bigg/ \sigma(y) 
  =  \frac{16 \pi}{3 M_F^3} \rho(y) \; , 
 \label{einstein}
\end{eqnarray}
where $\rho(y)= 1/2 (d \phi_0 (y) /dy)^2  + V(\phi_0)$ 
is the energy density of the bulk scalar field 
which generates the domain wall. 
Here, we assumed that the domain wall dominantly contributes to the energy 
momentum tenser, and thus the local behavior of the metric is controlled 
by the existence of the domain wall.  

Regarding the domain wall as the thin object like the D-brane, 
we discuss the effective theory on the brane based 
on the Nambu-Goto action $ S = - \int d^4x \Lambda^4 \sqrt{\hat{g}}$, 
by using the induced metric $\hat{g}(x, Y(x))$ \cite{sundrum}. 
The expansion of the determinant of the metric gives 
the action for the field $Y(x)$, which parameterizes the brane fluctuation, 
\begin{eqnarray}
  S =  \int d^4 x \; \Lambda^4 \; \sqrt{\bar g} 
   \left( \frac{1}{2} {\bar g}^{\mu \nu}  
   \partial_\mu Y(x)  \partial_\nu Y(x) 
   - \frac{1}{4} {\bar g}^{\mu \nu} {\bar g}_{\mu \nu , 5 5} Y(x)^2 
   + \cdots \right) \; ,  
 \label{induced}
\end{eqnarray}
where the index $5$ denotes the coordinate of the fifth dimension, 
and ${\bar g}_{\mu \nu} = g_{\mu \nu}(x, y=0)$ is the background metric 
at $y=0$. If the background metric is flat, $Y(x)$ is the massless state, 
which is just the NG boson. However, the non-trivial background metric 
seems to make the field massive. Using the metric assumed above 
and the Einstein equation of eq.(\ref{einstein}), 
the mass term of $Y(x)$ is found to be $m_Y^2 \sim \rho(0)/M_F^3$. 
From the results given in Sec. \ref{sec:set-up}, we can expect 
$ m_Y^2 \sim \Lambda^4 m_\phi/ M_F^3$, which is larger than $m_g^2$. 
Since $Y(x)$ gets the mass term, it seems to not be the NG boson itself. 
The would-be NG boson may be another field or a combination of $Y(x)$ 
and other fields. 

If there is another contribution to the energy momentum tensor, 
which is dominant compared with that of the domain wall, 
the mass term may be little related to the existence of the domain wall. 
In this case, $Y(x)$  may have much smaller mass or be massless 
and the would-be NG boson. In any case, to make a correct discussion, 
it seems to be essential to find a concrete solution of the extra dimensional 
theory with the domain wall configuration. Further work is needed. 

\acknowledgments
We would like to thank Yasuhiro Okada for useful discussions. 
This work was supported in part by the Grant-in-aid for Science and Culture
Research form the Ministry of Education, No.10740133 and "Priority
Area: Supersymmetry and Unified Theory of Elementary Particles (\#707)". 
N.O. is supported by the Japan Society for the Promotion of Science 
for Young Scientists.

%

%
\end{document}